\documentclass{revtex4}

\usepackage{graphicx}
\begin{document}
\bibliographystyle{revtex}


\title{Summary of Discussion Question 4: Energy Expandability of a Linear Collider}



\author{P.N. Burrows}
\email[]{p.burrows@physics.ox.ac.uk}
\affiliation{Oxford University}
\author{J.R. Patterson}
\email[]{ritchie@mail.lns.cornell.edu}
\affiliation{Cornell University}


\date{\today}

\begin{abstract}
We report on Discussion Question 4, in Sub-group 1 (`TeV-class')
of the Snowmass Working Group E3: `Experimental Approaches:
Linear Colliders', which addresses the energy expandability of a 
linear collider.  We first synthesize discussions of the energy reach
of the hardware of the 500 GeV designs for TESLA and NLC/JLC. Next,
we review plans for increasing the energy to 800-1000 GeV. We then
look at options for expanding the energies to 1500 GeV
and sketch the two-beam accelerator approach to achieving
multi-TeV energies. 
\end{abstract}

\maketitle

\section{Introduction}

The original statement of E3 SG1 DQ4 was:

`What is the energy for the initial phase of the LC (350
GeV, 500 GeV)? Assuming the energy reach is
expanded using the same technology as in the initial phase, what is the high energy
goal for a secondary phase of the LC? 
What R\&D issues remain before such an expansion is possible? What constraints on the
initial phase would ultimate conversion to
two-beam drive impose?' 

The scope of these questions, as posed, seemed too far-reaching for a
third-order sub-group to address. The questions of desirability
of a particular initial energy, or the energy goal for a secondary phase,
will clearly be determined by the physics that is (hopefully) discovered
between now and the Linear Collider turn-on, or upgrade, respectively. 
We feel that a reasonable summary of the current situation, in order of
decreasing confidence, is:

\begin{itemize}

\item
A minimum energy of 350 GeV will be needed to produce and 
study the top quark.

\item
Precision electroweak data imply that an energy of
500 GeV will be more than sufficient to produce the Standard Model (SM)
Higgs boson, if it exists, in association with the $Z^0$. If the 
95\% c.l. limit is taken seriously an energy of 300 GeV is sufficient.

\item
In all currently known SUSY models, the lightest SUSY Higgs boson lies
below 200 GeV in mass, and could therefore be produced by a collider with
energy of 300 GeV or more.

\item
Many, but not all, SUSY models predict new particles that can be 
pair-produced by a collider with energy 1000 GeV; many, but fewer, SUSY
models predict observable states at an energy of 500 GeV.
 
\item
Other beyond-SM models predict observable effects at $e^+e^-$ colliders
of energy beyond that explored at LEP2.

\end{itemize}

Discussion of particular models can be found elsewhere in these proceedings.

It therefore seems clear to us that it would be desirable to build a collider with 
the flexibility to operate initially within a range of energies
between 200 and 500 GeV. Physics
discoveries between now and its turn-on will serve to define the initial
energy, as well as the details of the optimal luminosity/energy 
running strategy during the first years of operation~\cite{grannis}.
It also seems clear to us that, almost independent of the details of
such discoveries before or at the collider, it is desirable that the
capability for an upgrade to the highest energy allowed by the technology
be designed from the start. Here we consider possible target energies for such an
upgrade of 1000 GeV and 1500 GeV.

We therefore modified the terms of E3 SG1 DQ4 to address the following
narrower set of questions:

\begin{enumerate}

\item
What is the maximum energy that could be achieved using ONLY
the hardware installed for the 500 GeV baseline design?
How would you achieve this; what is the corresponding luminosity?

\item
What would you need to do to achieve a c.m. energy of 1000 GeV:
a) at your `preferred' site, i.e. the one in your back yard, and 
(b) at an `ideal' site, i.e. one unconstrained by local geographical 
boundary conditions? What luminosity could you achieve? Are any changes
to the beam delivery system required? What are the incremental construction
and operating costs? What R\&D is needed?

\item
Same as the previous question, but relating to a c.m. energy of 1500 GeV.

\item
What do you need to do in your baseline design to allow eventual upgrade
to a multi-TeV collider?

\end{enumerate}

We posed these questions to Reinhard Brinkmann, Nobu Toge and Tor
Raubenheimer, accelerator physicists associated
with the TESLA~\cite{tesla} JLC~\cite{jlc} and NLC~\cite{nlc} projects
respectively.
We also received valuable input on the last of these questions from
Jean-Pierre Delahaye of the CLIC project~\cite{clic}, and from Ron Ruth~\cite{ruth}.
Here we attempt to synthesize the results of the presentations on these
issues
and the lively discussion that followed.

\section{The Maximum Energy Reach of the Nominal 500 GeV Collider}

We asked the machine designers to try to extrapolate the energy reach
of the nominal 500 GeV baseline machine under the assumption that they
received no funds for hardware upgrades. We also asked them to
evaluate the corresponding luminosity. 

\subsection{JLC/NLC}

\begin{figure}
 \includegraphics[width=80mm]{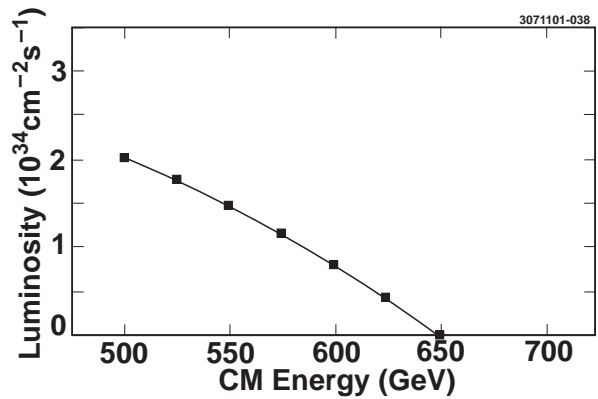}
 \caption{The projected NLC luminosity as a function of c.m. energy 
using only the hardware of the 500 GeV baseline design.  Energies above
500 GeV are attained by reducing the beam current and therefore the
beam-loading of the accelerating structures.}
 \label{NLClumvse}
 \end{figure}

For the X-band JLC/NLC the effective accelerating gradient in the linac is less than the
nominal value due to the effect of `beam loading.' The designers aim to
achieve a nominal unloaded mean gradient of 70 MV/m, corresponding to roughly
50 MV/m of actual mean gradient at nominal beam current of 190 bunches 
$\times$ $10^{10}$ particles/bunch. A tradeoff between beam current, i.e. luminosity,
and gradient, i.e. energy, is therefore possible. The maximum gradient of 70 MV/m, 
and hence c.m. energy of 650 GeV, is in principle 
achievable in the limit of zero current. 
Figure~\ref{NLClumvse} shows the expected luminosity vs. c.m. energy. 
If the energy were increased from 500 GeV to 600 GeV
the luminosity would decrease from 2 to 0.8 $\times$ $10^{34}$
cm$^{-2}$s$^{-1}$ 

\subsection{TESLA}

\begin{figure}
 \includegraphics[width=80mm]{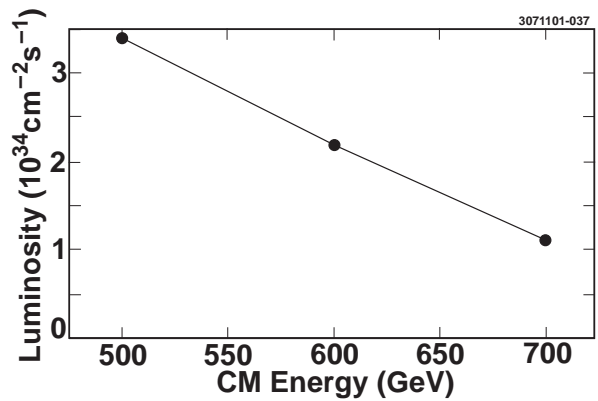}
 \caption{The projected TESLA luminosity as a function of c.m. energy 
using only the hardware of the 500 GeV baseline design.  Energies above
500 GeV are attained by increasing the gradient and therefore the
demands on the cooling plant.}
 \label{TESLAlumvse}
 \end{figure}

For TESLA, the accelerating gradient needed for 500 GeV c.m. energy is
23.4 MV/m. The cavities can be run at higher gradient, 
in principle up to and beyond 40 MV/m,
at the cost of higher power dissipation.   Hence, the headroom in the cryogenic
plant can be traded off to  achieve higher gradients.  
Using all of the 50\% built-in cooling overhead of the 500 GeV machine, 
it would in principle be possible to achieve a c.m. energy of 750 GeV,
however at reduced repetition rate below the nominal 5 Hz.
In order to contain the RF power at higher gradients, one must lower
the beam current (via increased bunch spacing) and
shorten the RF pulse length.  
Improved emittance
compensates for this loss of current so that with no other effects,
the luminosity would be unchanged.
In fact, the luminosity declines because at the higher gradient, 
the cavity filling
time (that is, the time required to achieve gradient) increases, 
so less of each pulse is useable for bunch acceleration.  
Figure~\ref{TESLAlumvse} shows the expected luminosity vs. c.m. energy. 
If the energy were increased from 500 GeV to 700 GeV
the luminosity would decrease from 3.4 to 1.1 $\times$ $10^{34}$
cm$^{-2}$s$^{-1}$.

\section{Energy upgradeability to 1000 GeV}

We asked the machine designers to explain the scenaria by which the
initial nominal energy of 500 GeV could be doubled. 

\subsection{JLC/NLC}
 
If the mean accelerating gradient for the X-band technology is assumed
to be fixed at a (loaded) value of 50 MV/m, the energy could be doubled
by doubling the RF system, i.e. doubling the number
of: accelerating structures, klystrons, modulators and power supplies.
This is anticipated in the JLC/NLC designs, the baseline 500 GeV site being designed
to accommodate the extra components without any need to lengthen the tunnels. 
Some increase in the cooling plant
would also be required. The A.C. power is estimated to rise by roughly 100 MW,
bringing the two-linac power to about 230 MW.
For c.m.  energies above 700 GeV the currently-envisioned
permanent magnets in the final doublet system of the final focus would need to
be replaced.
The total additional construction cost is estimated to be +25\%, or \$1.5B.

Key areas of R\&D for NLC are in the RF system and accelerating structures.
Since 1000 GeV is achieved by doubling the number of components, 
the R\&D for the 500 GeV baseline design also addresses 1000 GeV operation.
R\&D is ongoing at SLAC to demonstrate the reliable operation of Cu accelerating 
structures at high gradient. As of the time of writing,
travelling-wave cavities have been successfully operated with unloaded gradients
as high as 80 MV/m; it is desirable to reach 90 MV/m in order to allow more 
`headroom'.  This will need to be achieved for structures with the NLC
iris radius and with long-range wakefield suppression. 
The needed gradients
have been achieved in standing-wave cavities, which provide a
possible alternative, albeit one requiring substantial modification of
the linac design.

For the RF system, a full-scale 500 kV modulator is being tested,
an X-band klystron has been operated at low repetition rates, and 
low power operation of a delay line distribution system
has been demonstrated at the ATF in Japan.
A full-power test of a complete prototype RF unit is planned in 
the NLCTA in 2003.

The luminosity achievable at 1000 GeV is estimated to be 3.4 $\times$ $10^{34}$
cm$^{-2}$s$^{-1}$.

\subsection{TESLA}

The official goal of TESLA, as outlined in the TDR, is to reach a maximum
energy of 800 GeV. It is planned to achieve this by increasing the mean gradient
from 23 to 35 MV/m.  The cost to upgrade from 500 GeV to 800 GeV is
estimated to be 600-650 MEuro (an additional 20\% over the cost of the
500 GeV machine), which arises from the necessary expansion of the RF-power 
and cooling plants. The two-linac power consumption would increase to about 160 MW.

An ongoing R\&D program into the electropolishing of small
Nb cavities has yielded cavities that exceed 35 MV/m gradient.  These cavities
have yet to be tested under full power at the TTF.  It also remains
to test new cryogenic ``superstructures'', which improve the
packing-density of the cavities by 6\% by housing them in pairs, and 
therefore reduce the linac cost.  
 Tests of the superstructures will begin in 2002, leading
to a full scale test at TTF, possibly with the high-gradient cavities, in about 2004.

Theoretically, the maximum gradient achievable with the Nb cavities is
about 50 MV/m; naively this would allow an energy of more than 1 TeV to be reached,
but this is not thought to be realistic. Assuming a highest mean gradient of
35 MV/m, it would be necessary to 
lengthen one or both linacs in order to go beyond 800 GeV. 
At the DESY site two possibilities are:

a) extending the northern (positron) linac by 8 km; this would yield
asymmetric collisions of 625 GeV positrons on 400 GeV electrons. The cost
is estimated to be an additional 900MEuro, on top of the cost of the
800 GeV machine. 

b) adding 4 km to each linac to achieve symmetric 
500 $\oplus$ 500 GeV collisions. The incremental cost, on top of the 800 GeV
machine, would be similar to a), 
However, because of
the constraint imposed at the southern end by the River Elbe, the central site containing
the beam delivery system, the interaction region(s) and the detector(s)
would effectively have to be translated to the north to accommodate the extra 4 km
of southern linac.
This would have a serious impact on the present site planning, the
consequences of which would have to be carefully investigated.
The linac power load would increase by about 40MW to a total of 200 MW.
The beam delivery system would also have to be modified from its current
design to accommodate the
higher beam energy(ies); this might be achieved within the length of the present
design by applying a Raimondi-Seryi
scheme, as planned for the JLC/NLC.  

The luminosity achievable at 1000 GeV is estimated to be 6 $\times$ $10^{34}$
cm$^{-2}$s$^{-1}$.

\subsection{C-band JLC}

An alternative scheme for a 500 GeV collider has been developed in Japan; it is
based on C-band RF technology with a mean loaded gradient of 36 MV/m needed
in order to reach an energy of 500 GeV. 1000 GeV could be achieved either:

a) by doubling the active length of the RF system.

b) by building a site-filler tunnel from the start, 
increasing the gradient from 36 MV/m to 
50 MV/m, and adding RF as needed. The peak klystron power would need 
to be doubled from 50 MW to 100 MW; this, in addition to 
corresponding improvements in the other RF
components, would require an R\&D programme. Note that the gradient would
then be comparable with the X-band case.

\section{Energy upgradeability to 1500 GeV}

Although energy upgrades to 1000 GeV (JLC/NLC) and 800 GeV (TESLA) are
being designed into the projects, the foundations that would allow
eventual upgrade to energies substantially above 1000 GeV
seem less secure. Presumably an upgrade path, if possible,
will become clearer after many years of experience with either the X-band
or superconducting technology.

At the higher energies, AC power usage limits the beam current.  
TESLA and NLC therefore assume constant luminosity as their
energies increase from 1000 to 1500 GeV.  The high energy also increases
beamsstrahlung, and therefore energy spread, for both machines.

\subsection{JLC/NLC}
                                   
Assuming the use of the planned X-band technology,
two possibilities could be envisaged to achieve a c.m. energy of 1500 GeV:

a) If the accelerating gradient were fixed, at 50 MV/m loaded, an energy increase
would require a lengthening of the linacs. In the JLC/NLC design the injection
and pre-acceleration systems form a `trombone' structure at the head of each linac;
the trombones could be extended to accommodate longer linacs, with the addition
of transport lines.  This path would require additional klystrons.

b) If the site length were fixed, an energy increase would require an
increase in gradient, from 50 to 75 MV loaded. This would double the power
consumed by the RF system. The power could be preserved at about the same level
as that needed for 1 TeV operation by running at a repetition rate of
90 Hz rather than the nominal 120 Hz; this would imply a luminosity reduction
of roughly 25\%.

\subsection{TESLA}
                                   
Assuming the use of the planned superconducting technology, with a maximum
mean gradient of 35 MV/m, the linacs would need to be lengthened by 
a total of 14 km in order to
achieve a c.m. energy of 1500 GeV.  The total cost of such a machine would be
in the ballpark of 5.5 to 6 BEuro.
Given geographical limitations at DESY, either
the central site would have to be relocated, as discussed in the previous
section, or the tunnel would need to be dug deeper from the project start,
so as to allow tunnel passage beneath the River Elbe. 
The latter is not currently envisaged.

\section{Energy upgradeability to multi-TeV scales}

Achieving very high energies requires both very high accelerating gradients and
very efficient delivery of RF power.  Even in an ideal case, it is likely that the
luminosity achieved will be limited by the need to contain power use.

The most mature ideas for a Multi-TeV $e^+ e^-$ collider are based on so-called
`two-beam' technology. This is the basis of the design for CLIC~\cite{clic,delahaye}, which is 
under development at CERN. In the CLIC design, power is `stored' in a high-current (190A),
 low
energy (1.3 GeV) drive beam and transferred to the low-current, high-energy main beam.  
The design requires an accelerating gradient of 150 MV/m in
Cu structures operating at 30 GHz;
gradients above 100 MV/m have been 
achieved, but so far only in single-cell structures operating at low power. 
In order to achieve a useful luminosity, CLIC will need very small 
emittance and beam spots; in the vertical plane the beam size must be less
than 1 nanometre, a factor of 3 and 5 smaller than the already-ambitious
beam sizes planned for JLC/NLC and TESLA, respectively.
 In order to achieve this, the emittance must be preserved in the face of vibrations, 
strong wakefields and other disruptions. Like the NLC, CLIC requires
a crossing angle (20 mrad) in order to avoid parasitic collisions by
closely-spaced bunches.
CERN is now planning a  facility, CTF3, that will test the two-beam accelerator
proof-of-principle.  The facility will also test whether 
gradients of 150 MV/m can be sustained under operating conditions. 
The goal of the designers is to have demonstrated the
needed technology by 2008. 

Should the CLIC design prove feasible, it will draw on experience gained at a 
first generation linear collider, particularly in the area of attaining and preserving low
 emittance.  Many of these extreme challenges are shared by the
NLC/JLC, though CLIC would push them substantially further.

A scheme was presented~\cite{ruth} for an `adiabatic' upgrade of the 
JLC/NLC to multi-TeV operation via phased installation of a drive beam, 
doubling of the RF frequency, and replacement of the accelerating structures.
This plan for multi-TeV operation is similar to CLIC's, and shares
critical R\&D areas.


\section{Summary and Conclusions}
Using the hardware envisioned for 500 GeV operation at TESLA and NLC/JLC,
it should be possible to push the c.m. energy of a linear collider
up to about 650 GeV at the cost of reduced luminosity.
Expansion to energies in the 800-1000 GeV range, at full luminosity,
is incorporated into the designs of both TESLA and NLC/JLC.
Reaching c.m. energies above 
1000 GeV goes beyond the current plans, 
and if such energies are attainable, achieving them will 
build on experience gained at 1000 GeV and
below.  The multi-TeV regime requires a new strategy
such as the two-beam technique envisioned for CLIC.  The parameters
for such a machine might be established late this decade.

\begin{acknowledgments}
We are grateful to:
Gary Bower, 
Reinhard Brinkman, 
Jean-Pierre Delahaye, 
Gilbert Guignard,
Tor Raubenheimer, 
Ron Ruth,
Nobu Toge,
Nick Walker,
to many more for lively contributions to the discussion,
and to countless others
 for their patience, forbearance and good humour during our proceedings.
\end{acknowledgments}

\end{document}